\documentclass[aps,prc,floatfix,superscriptaddress,a4paper,
               nofootinbib,preprint]{revtex4}

\usepackage{graphicx}
\usepackage{color}

\usepackage{amsmath}
\usepackage{amssymb}
\usepackage{enumerate}

\newcommand{\be}{\begin{equation}}
\newcommand{\ee}{\end{equation}}

\usepackage{bm}

\usepackage{color}

\def \red#1 {{\color{red}#1}}

\newcommand{\eq}[1]{\begin{align} #1 \end{align}}
\newcommand{\p}{\!+\!}

\usepackage{setspace}

\usepackage{algorithm}
\usepackage{algorithmicx}
\usepackage{algpseudocode}

\algnewcommand\algorithmicforeach{\textbf{for each}}
\algdef{S}[FOR]{ForEach}[1]{\algorithmicforeach\ #1\ \algorithmicdo}
\RequirePackage{tikz}
\RequirePackage{zref-abspage}
\RequirePackage{zref-user}
\RequirePackage{tikz}
\RequirePackage{atbegshi}
\usetikzlibrary{calc}
\RequirePackage{tikzpagenodes}
\RequirePackage{etoolbox}

\makeatletter
\newcommand*\ALG@lastblockb{b}
\newcommand*\ALG@lastblocke{e}
\apptocmd{\ALG@beginblock}{%
	\ifx\ALG@lastblock\ALG@lastblockb
	\ifnum\theALG@nested>1\relax\expandafter\@firstoftwo\else\expandafter\@secondoftwo\fi{\ALG@tikzborder}{}%
	\fi
	\let\ALG@lastblock\ALG@lastblockb%
}{}{\errmessage{failed to patch}}

\pretocmd{\ALG@endblock}{%
	\ifx\ALG@lastblock\ALG@lastblocke
	\addtocounter{ALG@nested}{1}%
	\addtolength\ALG@tlm{\csname ALG@ind@\theALG@nested\endcsname}%
	\ifnum\theALG@nested>1\relax\expandafter\@firstoftwo\else\expandafter\@secondoftwo\fi{\endALG@tikzborder}{}%
	\addtolength\ALG@tlm{-\csname ALG@ind@\theALG@nested\endcsname}%
	\addtocounter{ALG@nested}{-1}%
	\fi
	\let\ALG@lastblock\ALG@lastblocke%
}{}{\errmessage{failed to patch}}

\tikzset{ALG@tikzborder/.style={line width=0.5pt,black}}

\newcommand*\currenttextarea{current page text area}

\newcommand*{\updatecurrenttextarea}{%
	\if@twocolumn
	\if@firstcolumn
	\renewcommand*{\currenttextarea}{current page column 1 area}%
	\else
	\renewcommand*{\currenttextarea}{current page column 2 area}%
	\fi
	\else
	\renewcommand*\currenttextarea{current page text area}%
	\fi
}

\newcounter{ALG@tikzborder}
\newcounter{ALG@totaltikzborder}
\newenvironment{ALG@tikzborder}[1][]{%
	\ifx&#1&\else
	\tikzset{ALG@tikzborder/.style={#1}}%
	\fi
	\stepcounter{ALG@totaltikzborder}%
	\expandafter\edef\csname ALG@ind@border@\theALG@nested\endcsname{\theALG@totaltikzborder}%
	\setcounter{ALG@tikzborder}{\csname ALG@ind@border@\theALG@nested\endcsname}%
	\tikz[overlay,remember picture] \coordinate (ALG@tikzborder-\theALG@tikzborder);% node {\theALG@tikzborder};% Modified \tikzmark macro
	\zlabel{ALG@tikzborder-begin-\theALG@tikzborder}%
	\ifnum\zref@extract{ALG@tikzborder-begin-\theALG@tikzborder}{abspage}=\zref@extract{ALG@tikzborder-end-\theALG@tikzborder}{abspage} \else
	\updatecurrenttextarea
	\ALG@drawvline{[shift={(0pt,.5\ht\strutbox)}]ALG@tikzborder-\theALG@tikzborder}{\currenttextarea.south east}{\ALG@thistlm}%
	\newcounter{ALG@tikzborderpages\theALG@tikzborder}%
	\setcounter{ALG@tikzborderpages\theALG@tikzborder}{\numexpr-\zref@extract{ALG@tikzborder-begin-\theALG@tikzborder}{abspage}+\zref@extract{ALG@tikzborder-end-\theALG@tikzborder}{abspage}}%
	\ifnum\value{ALG@tikzborderpages\theALG@tikzborder}>1
	\edef\nextcmd{\noexpand\AtBeginShipoutNext{\noexpand\ALG@tikzborderpage{\theALG@tikzborder}{\the\ALG@thistlm}}}%some pages need a border on the whole page
	\nextcmd
	\fi
	\fi
}{%
\setcounter{ALG@tikzborder}{\csname ALG@ind@border@\theALG@nested\endcsname}%
\tikz[overlay,remember picture] \coordinate (ALG@tikzborder-end-\theALG@tikzborder);% node {\theALG@tikzborder};% Modified \tikzmark macro
\zlabel{ALG@tikzborder-end-\theALG@tikzborder}%
\updatecurrenttextarea
\ifnum\zref@extract{ALG@tikzborder-begin-\theALG@tikzborder}{abspage}=\zref@extract{ALG@tikzborder-end-\theALG@tikzborder}{abspage}\relax
\ALG@drawvline{[shift={(0pt,.5\ht\strutbox)}]ALG@tikzborder-\theALG@tikzborder}{ALG@tikzborder-end-\theALG@tikzborder}{\ALG@thistlm}%
\else
\ALG@drawvline{\currenttextarea.north west}{ALG@tikzborder-end-\theALG@tikzborder}{\ALG@thistlm}%
\fi
}

\newcommand*{\ALG@drawvline}[3]{%#1=from, #2=to, #3=value of \ALG@tlm/\ALG@thisthm
	\begin{tikzpicture}[overlay,remember picture]
	\draw [ALG@tikzborder]
	let \p0 = (\currenttextarea.north west), \p1=(#1), \p2 = (#2)
	in
	(#3+\fboxsep+14\pgflinewidth+\x0,\y1+\fboxsep+.5\pgflinewidth)%-\fboxsep-.5\pgflinewidth
	--
	(#3+\fboxsep+14\pgflinewidth+\x0,\y2-\fboxsep-.5\pgflinewidth)
	;
	\end{tikzpicture}%
}

\newcommand{\ALG@tikzborderpage}[2]{%the whole page gets a border, #1=value of \theALG@tikzborder, #2=value of \ALG@tlm/\ALG@thistlm
	\updatecurrenttextarea
	\setcounter{ALG@tikzborder}{#1}%
	\ALG@drawvline{\currenttextarea.north west}{\currenttextarea.south east}{#2}%
	\addtocounter{ALG@tikzborderpages\theALG@tikzborder}{-1}%
	\ifnum\value{ALG@tikzborderpages\theALG@tikzborder}>1
	\AtBeginShipoutNext{\ALG@tikzborderpage{#1}{#2}}%
	\fi
	\vspace{-0.5\baselineskip}% Compensate for the generated extra space at begin of the page. No idea why exactly this happens.
}

\def\ALG@tikzbordertext{\the\ALG@tlm}
\renewcommand{\ALG@beginalgorithmic}{\scriptsize}
\algrenewcommand\alglinenumber[1]{\tiny #1:}
\makeatother
\begin{document}

\title{Steady state of isolated systems versus microcanonical ensemble in 
cell model of particle creation and
annihilation}

\author{M. Gazdzicki}
\affiliation{Geothe-University Frankfurt am Main, Germany}
\affiliation{Jan Kochanowski University, Kielce, Poland}

\author{M. I. Gorenstein}
\affiliation{Bogolyubov Institute for Theoretical Physics, Kiev, Ukraine}
\affiliation{Frankfurt Institute for Advanced Studies, Frankfurt, Germany}

\author{A. Fronczak}
\affiliation{Faculty of Physics, Warsaw University of Technology, Warsaw, Poland}

\author{P. Fronczak}
\affiliation{Faculty of Physics, Warsaw University of Technology, Warsaw, Poland}

\author{M. Mackowiak-Pawlowska}
\affiliation{Faculty of Physics, Warsaw University of Technology, Warsaw, Poland}

\begin{abstract}
A simple model of particle creation and annihilation in an isolated assembly of particles with conserved energy and fixed volume, the Cell Model, is formulated. With increasing time, particle number distribution, obtained by averaging over many systems, approaches a time-independent, steady state distribution. Dependence of the steady state distribution on creation and annihilation conditional reaction probabilities is studied.
The results obtained for the steady state are compared with predictions of statistical mechanics 
within the microcanonical ensemble. In general, 
the predictions of both models are different. 
They agree only if the creation and annihilation conditional probabilities are equal.
This condition also results in the detailed balance in the steady state.

\end{abstract}



\maketitle
\section{Introduction}
\label{sec:introduction}

Isolated systems of interacting particles are expected to evolve from an arbitrary starting conditions to a steady state (SS), i.e. the state whose macroscopic properties are independent of time. 
Numerous experimental results on properties of many-particle SSs are in agreement with statistical mechanics.

The basic ensemble of microstates in statistical mechanics is the one formulated for isolated systems~\cite{LL}, the so-called microcanonical ensemble (MCE). Other statistical ensembles, like canonical 
and grand canonical ensembles, can be straightforwardly derived from the MCE. 
The MCE is defined by all conserved quantities of the system, its volume $V$ and possible microstates. 
Here, it is assumed that energy $E$ is the only conserved quantity. Thus, there are only two independent thermodynamic variables, $E$ and $V$.

The key postulate of statistical mechanics is that all MCE microstates have equal probabilities to appear. 
Properties of the MCE, like particle number distribution, are easy to calculate for an ideal gas of particles in which 
the total energy is equal to the sum of single particle energies. Obviously, the energy conservation introduces a correlation between particles.

In contrast to statistical mechanics, dynamical models do not postulate  probabilities of the microstates. 
Instead, these probabilities depend on properties of elastic and inelastic reactions between particles. 
The reactions lead to a time walk of a dynamical system from microstate to microstate starting  
from an assumed microstate at $t=0$.
Considering a large number of systems with $E$ and $V$ at a given time $t$ 
one gets an ensemble of microstates. In general, the microstate probability distribution depends on time 
and a starting microstate distribution. Let us assume that the probability distribution 
for $t\rightarrow\infty$ approaches a distribution which is independent of a starting distribution of microstates. 
This asymptotic distribution will be called the SS distribution, and the corresponding ensemble of microstates 
will be referred to as the SS ensemble. One expects that the SS depends on properties of reactions between particles. They determine
probabilities of transitions between microstates. Of course, the MCE lacks such a dependence. Thus, in general, 
the SS ensemble is expected to differ from the MCE. 

Probably the closest to isolated systems are multi-particle systems created in high-energy collisions between particles or nuclei.
Surprisingly, they show many features which can be described within
statistical mechanics~\cite{Becattini:1997rv,Becattini:2003wp,Andronic:2005yp}.
In particular, MCE well describes basic properties of experimental
data~\cite{Becattini:1995if,Ferroni:2011fh,Begun:2006uu}. 
This can be modelled by postulating that each collision leads to 
instantaneous creation of a microstate randomly selected from the 
MCE~\cite{Hagedorn:1965st,Gazdzicki:1998vd,Castorina:2007eb}.
On the other hand, popular dynamical models of multi-particle production postulate a set of allowed inelastic reactions and assume  
their conditional probabilities. By closing particles in an isolated box, one can calculate properties of the SS of a given dynamical model.
For example, this was done for the relativistic quantum molecular dynamics model~\cite{belkacem}). The conclusion was that the SS differs from the corresponding MCE.

We did not find in the literature any discussion of the difference between the
predictions of statistical mechanics and SSs calculated within the dynamical
models of multi-particle production, even though statistical and dynamical 
approaches are very popular in modelling high energy collisions.

In this paper, we introduce a simple approach for modelling an assembly of identical particles which are subject to creation and annihilation reactions, the Cell Model (CM).
The important feature of the model is that it characterizes the time dependence of the 
particle assembly as due to stochastic processes, rather than under deterministic and time-reversal invariant Hamiltonian dynamics.
Correlations due to quantum statistics are not considered. The CM satisfies the energy conservation and allows to calculate the particle number distribution. 
Conditions needed to reach the agreement between the SS and MCE predictions 
are found and discussed.

The paper is organized as follows. Section~\ref{sec:cm} introduces the CM. 
In Sec.~\ref{sec:mce}, the MCE corresponding to the CM is formulated. The model presented here is a Markov chain on a countable microstate space that follows a random walk. This allows to calculate a time-dependence of particle number distribution using a Monte Carlo technique as presented in Sec.~\ref{sec:cell}. Moreover analytical results for the SS are obtained in Sec.~\ref{sec:MEq} by solving the corresponding master equation (MEq).
Section~\ref{sec:SSvsMCE} presents a comparison between the results for the SS and MCE. 
The conditions needed for the agreement between the SS and MCE are derived and discussed. Closing remarks presented in Sec.~\ref{sec:sum} at the end of the paper.

\section{Cell Model: Basic Assumptions}
\label{sec:cm}

Markov chain models and their approximation via the MEq are reviewed  
in the context of statistical mechanics in Section on {\it Stochastic dynamics} of 
{\it Compendium of the foundations of classical statistical physics}~\cite{Uffink}.

To assure minimum confusion in subsequent sections, we first
summarize the terminology employed here:
\begin{enumerate}[(i)]
\item
The label {\it system} refers to an assembly of particles with energy $E$
and volume $V$. Number of particles $N$ in the system changes in time $t$
due to reaction between particles.
\item
{\it Microstate} is a specific microscopic configuration of particles in the system at time $t$.
\item
{\it State} refers to an ensemble of microstates.
\item
{\it SS} labels an ensemble of microstates to which systems
approach for $t\longrightarrow\infty$.
\item
{\it Ergodicity} refers to a model property that any microstate has zero probability to never recur.
\item
{\it Particle number distribution}, $P(N,t)$, is a probability distribution of $N$ calculated for many systems at time $t$. Mean particle number is given by $\langle N \rangle(t)~\equiv~\sum\limits_N N\,P(N,t)$.
\end{enumerate}

The CM is introduced to study properties of isolated systems in the SS and compare the results with the corresponding MCE findings. In this paper, the comparison is limited to particle number distribution. The number of particles in the system changes in time due to creation and annihilation reactions (inelastic reactions) which take place in $V$ space cells\footnote{The number of cells  in the CM plays the role of volume in statistical mechanics.}, $v=1,\;\ldots,\;V$. Then, a new distribution of particles in cells is drawn from all possible microstates with a given particle number. This mimics the effect of elastic reactions. The formation time of a new microstate is assumed to be $\Delta t=1$. Thus, in the model, time is discrete, $t=0,1,2,\dots~$.

It is assumed that reactions are taking place independently in cells, and energy is conserved in each reaction separately. This local energy conservation leads, of course, to the energy conservation in the whole system.
For simplicity, only two values of single particle energy are allowed:
$\varepsilon=1$ and $2$. The microstates in the CM are thus defined by the set of ''occupation numbers'' $(n_{1,1},n_{2,1};\, \ldots \,; n_{1,V},n_{2,V})$, where $n_{\varepsilon,v}$ gives the number of particles having energy $\varepsilon$ in the $v$-th cell. All possible combinations of these sets should satisfy the energy conservation,
\eq{\label{en-cons}
\sum_{v=1}^Vn_{1,v}+\sum_{v=1}^V 2n_{2,v}~\equiv ~N_1+2\,N_2~=~E~.
}

%____________________________________________________________________MonteCarlo
\section{Cell Model: Microcanonical Ensemble}\label{sec:mce}

In this section, the MCE which corresponds to the CM is formulated
and the MCE particle number distribution is derived.
For given $E$ and  $N=N_1+N_2$, one gets
% $N_1$ and $N_2$ as:
\begin{equation}\label{CMEdefN1N2}
N_1=\sum_{v=1}^Vn_{1,v}=2\,N-E~,\;\;\;\;\;\mbox{and}\;\;\;\;\;
N_2=\sum_{v=1}^Vn_{2,v}=E-N~.
\end{equation}
Consequently, the number of microstates is equal to
\begin{eqnarray}\label{CMEUWN}
W_{\rm mce}(N;E,V)\;=\;W(N_1;E,V)W(N_2;E,V)
\;=\;\binom{N_1+V-1}{N_1}\binom{N_2+V-1}{N_2}~,
\end{eqnarray}
i.e., for each arrangement of $N_1$ indistinguishable particles of energy $\varepsilon=1$ in $V$ cells, one can have any arrangement of $N_2$
indistinguishable particles of energy $\varepsilon=2$ in $V$ cells. The expressions for $W(N_\varepsilon;E,V)$ correspond to the 
well-known result~\cite{bookFeller} for the number of different arrangements of $N_\varepsilon$  unlabeled balls (particles) among $V$ labeled boxes (cells). The assumption of indistinguishable (unlabeled) particles implies that any permutation of particles, either in a cell or between different cells, does not produce
a new microstate.

The total number of microstates for all possible pairs $(N_1,N_2)$,
$W_{\rm mce}(E,V)$, is given by:
\begin{eqnarray}\label{CMEUWE}
W_{\rm mce}(E,V)\;=\sum_{N=\lceil\frac{E}{2}\rceil}^E W_{\rm mce}(N;E,V)~,
\end{eqnarray}
where
$\lceil\frac{E}{2}\rceil=\frac{E}{2}$ for even $E$, and
$\lceil\frac{E}{2}\rceil=\frac{E+1}{2}$ for odd $E$.

The MCE assumes that all microstates with energy $E$ (Eq.~(\ref{en-cons}))
appear with equal probability. This allows to calculate the MCE particle number distribution as:
\begin{equation}\label{CMEUdefPN}
P_{\rm mce}(N;E,V)~=~\frac{W_{\rm mce}(N;E,V)}{W_{\rm mce}(E,V)}~.
\end{equation}

Examples of $P_{\rm mce}(N;E,V)$ are shown in the following sections.
%%%%%%%%%%%%%%%%%%%%%%%%%%%%%%%%%%%%%%%%%%%%%%%%%%%%%%%%%%%%%%%%%

%____________________________________________________________________MonteCarlo
\section{Cell Model: Monte Carlo Approach}\label{sec:cell}

In this section, the time evolution of a system is considered using a discrete-time Monte Carlo approach. It starts at $t=0$ from an arbitrary microstate and then, in the consecutive time steps $t\geq 0$, it runs as follows:
\begin{enumerate}[(i)]

\item The initial numbers of particles with energy $\varepsilon=1$ and $2$, which are correspondingly equal to $N_1^{\rm I}(t)$ and $N_2^{\rm I}(t)$ (where $N_1^{\rm I}(t)+2N_2^{\rm I}(t)=E$), are used to randomly draw an initial microstate $(n^{\rm I}_{1,1},n^{\rm I}_{2,1};\;\ldots;\;n^{\rm I}_{1,V},n^{\rm I}_{2,V})$ with $\sum_{v=1}^Vn_{1,v}=N_1^{\rm I}$ and $\sum_{v=1}^Vn_{2,v}=N_2^{\rm I}$. This ''random drawing'' mimics the effect of elastic reactions.

\item Then, creation and annihilation reactions take place. The reactions change the initial microstate, $(n^{\rm I}_{1,1},n^{\rm I}_{2,1};\;\ldots;\;n^{\rm I}_{1,V},n^{\rm I}_{2,V})$, into a final one,
$(n^{\rm F}_{1,1},n^{\rm F}_{2,1};\;\ldots;\;n^{\rm F}_{1,V},n^{\rm F}_{2,V})$,
with particle numbers $N_1^{\rm F}(t)=\sum_{v=1}^Vn_{1,v}^{\rm F}$
and $N_2^{\rm F}(t)=\sum_{v=1}^Vn_{2,v}^{\rm F}$.

\item The multiplicities $N_1^{\rm F}(t)$ and $N_2^{\rm F}(t)$ are then used
to draw (elastic reactions) an
initial microstate at $t+1$ with
$N_\varepsilon ^{\rm I}(t+1)=N_\varepsilon ^{\rm F}(t)$.
\item
Then the above sequence of executing elastic
and inelastic reactions repeats in the next time step.
\end{enumerate}

The assumed local energy conservation, that is met in each cell, implies the following relation between the initial and final cell multiplicities:
\begin{equation}\label{localE}
n_{1,v}^{\rm I}(t)+2\,n_{2,v}^{\rm I}(t)~=~n_{1,v}^{\rm F}(t)+2\,n_{2,v}^{\rm F}(t)~,
\end{equation}
and the global energy conservation can be expressed as
\begin{equation}\label{defE}
N_1^{\rm F}(t)+2\,N^{\rm F}_2(t)~=~E~=~{\rm const}~.
\end{equation}
Correspondingly, the total number of particles at time $t$ is equal to
\begin{equation}\label{defN}
N(t)~=~N^{\rm F}_1(t)+N^{\rm F}_2(t)~.
\end{equation}
%%%%%%%%%%%%%%%%%%%%%%%%

Here, we consider a version of the CM in which the following creation and annihilation reactions $(n^I_{1,v},n^I_{2,v})\rightarrow (n^F_{1,v},n^F_{2,v})$
take place:
\begin{enumerate}[(i)]
\item Annihilation $(3,0) \rightarrow (1,1)$ with probability $A_3$,
\item Annihilation $(2,1) \rightarrow (0,2)$ with probability $A_4$,
\item Creation $(1,1) \rightarrow (3,0)$ with probability $C_3$,
\item Creation $(0,2) \rightarrow (2,1)$ with probability $C_4$.
\end{enumerate}
Indices 3 and 4 indicate the cell energy. The probabilities $A$ and $C$ are conditional probabilities of a reaction given an initial cell multiplicity. They are assumed to be independent of time and range from zero to one.

Then, for any values of $E$, $V$ and conditional reaction probabilities, the time evolution of the system can be calculated using a Monte Carlo simulation. Note that the implementation of the classical indistinguishable particles in dynamical simulations is not a straightforward task. Thus, we provided a practical algorithm for this problem (see Appendix~\ref{App1}).

\begin{figure*}[t]
	\begin{center}
		\includegraphics[width=0.95\textwidth]{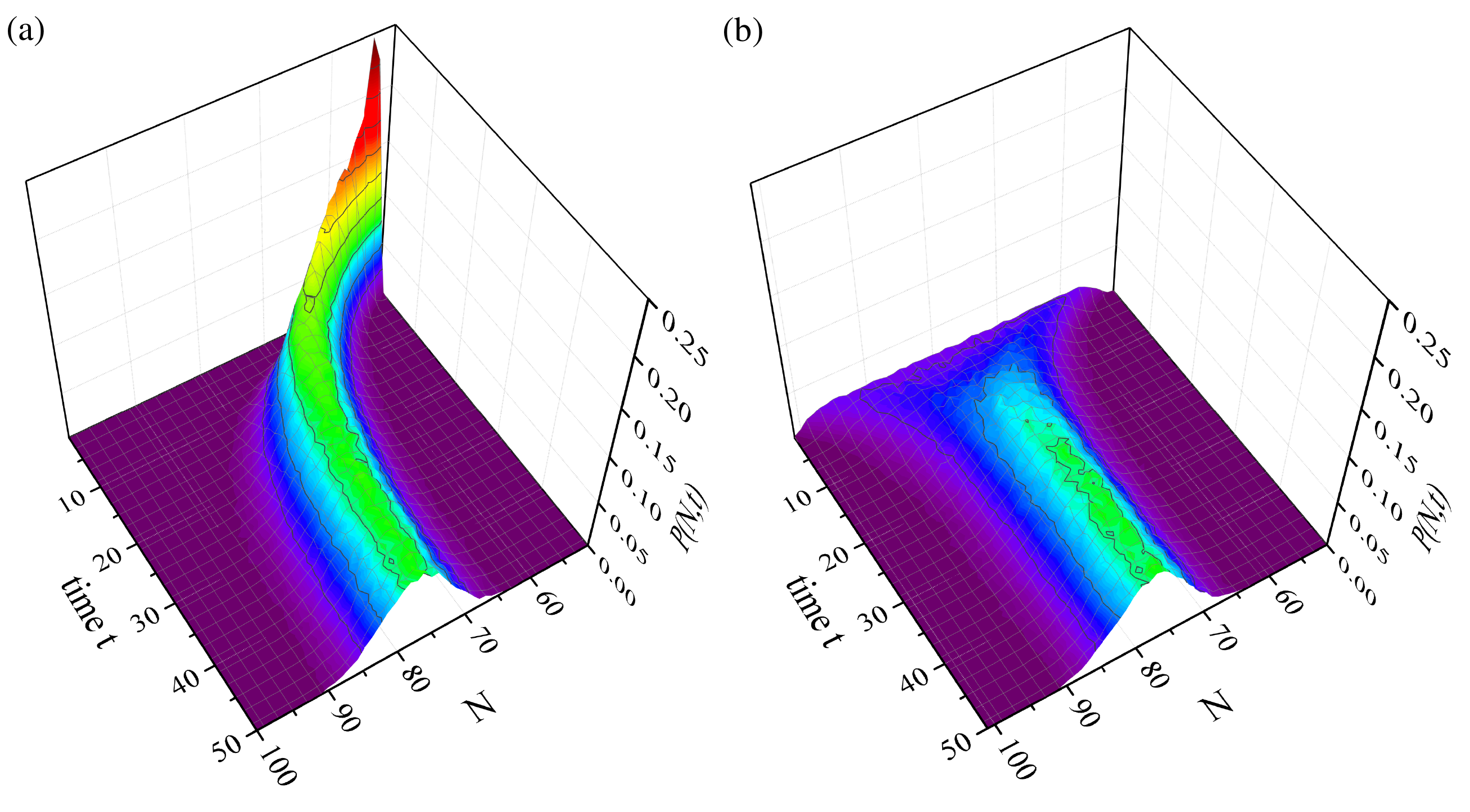}
	\end{center}
	\caption{Examples of time evolution of $P(N,t)$ for CM with $E=100$ and $V=20$. (a) $P(N,0)=\delta_{N,50}$, (b) $P(N,0)=const$.}
	\label{fig-time}
\end{figure*}

By construction, the particle number $N$ at time $t$ is a random variable. Thus, when following the time evolution of many systems starting from the same microstate, a non-trivial particle number distribution $P(N,t)$ is obtained\footnote{Note that since $N^{\rm F}_1(t)=N^{\rm I}_1(t+1)$ and $N^{\rm F}_2(t) = N^{\rm I}_2(t+1)$, it follows that for a system approaching a SS one has $P(N^{\rm F})=P(N^{\rm I})$.}. This is illustrated in Fig.~\ref{fig-time} (a), which shows time dependence of the multiplicity distribution for $E=100$, $V=20$ and the initial multiplicity distribution $P(N,0)=\delta_{N,50}$. In Fig.~\ref{fig-time} (b), the time evolution of $P(N,t)$ for the same system parameters but completely different initial microstate distribution, i.e. $P(N,0)=const$, is shown to converge to the same SS distribution like in the former case, i.e.
\begin{equation}\label{pom1}
\lim_{t\rightarrow\infty}P(N,t)=P_{\rm ss}(N;E,V).
\end{equation}
The above observation arises from the fact that the CM is an ergodic Markov chain, and as such, regardless of the initial conditions, it is always characterized by a unique stationary distribution (see Appendix~\ref{Markov}).

\begin{figure}[t]
	\begin{center}
		\includegraphics[width=0.95\textwidth]{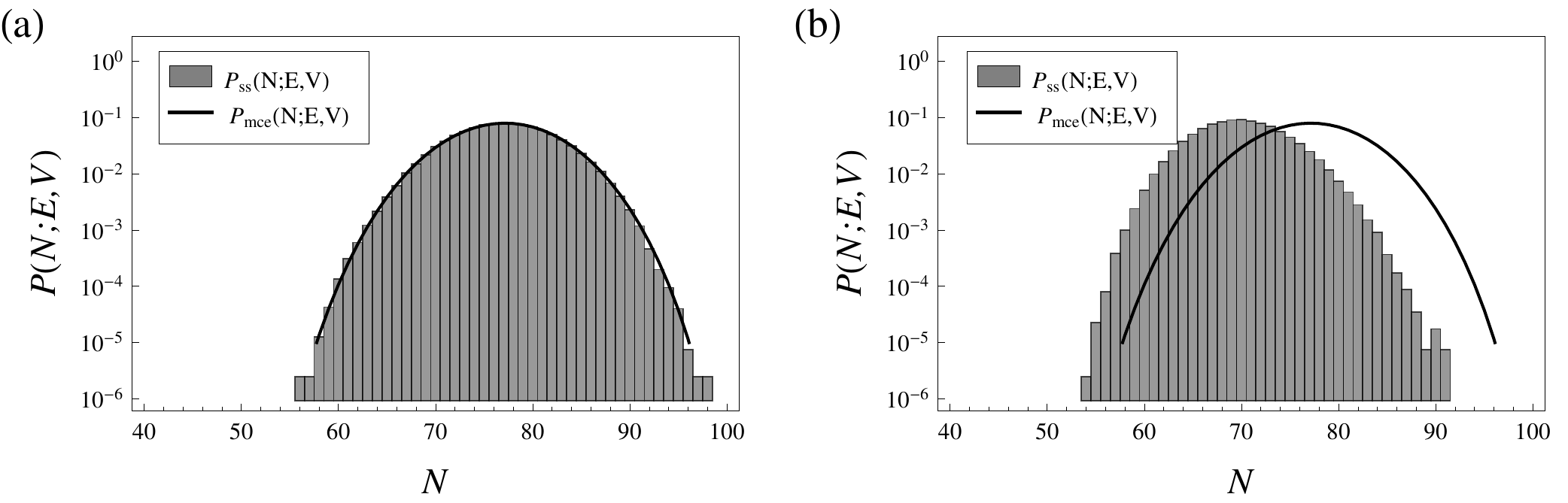}
	\end{center}
	\caption{The SS particle number distribution as compared to the MCE one. Gray histograms show numerical Monte Carlo simulation results for the CM  with $V=10$ and $E=100$. Lines represent the MCE results~(\ref{CMEUdefPN}). Conditional reaction probabilities used are the following: (a)~$A_3\!=\!A_4\!=\!C_3\!=\!C_4\!=\!1$, (b)~$A_3\!=\!A_4\!=\!C_4\!=\!1,\;C_3\!=\!\frac{1}{2}$.}
	\label{fig1}
\end{figure}

Figure~\ref{fig1} presents a comparison between $P_{\rm mce}(N;E,V)$ (\ref{CMEUdefPN}) and $P_{\rm ss}(N;E,V)$ (\ref{pom1}) for different settings of the model parameters. The agreement between the SS and MCE distributions is observed only
in some special cases, e.g., for equal conditional reaction probabilities: $A_3\!=\!A_4\!=\!C_3=\!C_4$. In the case of arbitrarily-set conditional reaction probabilities, $P_{\rm mce}(N;E,V)$ and $P_{\rm ss}(N;E,V)$
usually differ significantly.

\section{Cell Model: Analytical Approach}\label{sec:MEq}

In this section, an analytical expression for the particle number distribution in the SS is derived within the discrete-time MEq approach (see, e.g., Ref.~\cite{bookMEq}). It is assumed
that for a system in the SS a probability of having more
than one inelastic reaction at a given time step can be neglected.
This leads to the following MEq for the time dependence of
particle number probability distribution, $P(N,t)$:
\begin{eqnarray}
\nonumber\Delta P(N,t)=&A_3&\,
\frac{\binom{(N_1-1)+V-2}{V-2 } \binom{(N_2-1)+V-2}{V-2}}
{\binom{(N_1+2)+V-1}{V-1} \binom{(N_2-1)+V-1}{V-1}}\;P(N\!+\!1,t)
\\\nonumber
+&A_4&\,
\frac{\binom{N_1+V-2}{V-2} \binom{(N_2-2)+V-2}{V-2}}
{\binom{(N_1+2)+V-1}{V-1} \binom{(N_2-1)+V-1}{V-1}}
\;P(N\!+\!1,t)
\\\nonumber
+&C_3&\,
\frac{\binom{(N_1-3)+V-2}{V-2}\binom{N_2+V-2}{V-2}}
{\binom{(N_1-2)+V-1}{V-1} \binom{(N_2+1)+V-1}{V-1}}
\;P(N\!-\!1,t)
\\\nonumber
+&C_4&\,
\frac{\binom{(N_1-2)+V-2}{V-2}\binom{(N_2-1)+V-2}{V-2}}
{\binom{(N_1-2)+V-1}{V-1} \binom{(N_2+1)+V-1}{V-1}}
\;P(N\!-\!1,t)
\\\nonumber
-&A_3&\,
\frac{\binom{(N_1-3)+V-2}{V-2}\binom{N_2+V-2}{V-2}}
{\binom{N_1+V-1}{V-1}\binom{N_2+V-1}{V-1}}
\;P(N,t)
\\\nonumber
-&A_4&\,
\frac{\binom{(N_1-2)+V-2}{V-2} \binom{(N_2-1)+V-2}{V-2}}
{\binom{N_1+V-1}{V-1}\binom{N_2+V-1}{V-1}}
\;P(N,t)
\\\nonumber
-&C_3&\,
\frac{\binom{(N_1-1)+V-2}{V-2}\binom{(N_2-1)+V-2}{V-2}}
{\binom{N_1+V-1}{V-1}\binom{N_2+V-1}{V-1}}
\;P(N,t)
\\
-&C_4&\,
\frac{\binom{N_1+V-2}{V-2}\binom{(N_2-2)+V-2}{V-2}}
{\binom{N_1+V-1}{V-1}\binom{N_2+V-1}{V-1}}
\;P(N,t)~,
\label{meq}
\end{eqnarray}
where $\Delta P(N,t)=P(N,t+1)-P(N,t)$.

The first term in the right-hand side of Eq.~(\ref{meq}) states that systems with $N$ particles ($N_1$ with energy $1$ and $N_2$ with energy~$2$) are created by reducing by one particle multiplicity in systems with $N\!+\!1$  particles, as a result of the reaction: $(3,0)\rightarrow (1,1)$.
For this reaction, initial numbers of particles with energy 1 and 2 are
equal to $N_1\!+\!2$ and $N_2\!-\!1$, respectively. Probability of
this reaction is the product $w_1w_2w_3$ of
probabilities of three independent events, namely:
\begin{enumerate}[(i)]
\item $w_1=P(N\!+\!1,t)$ to have $N\!+\!1$ particles ($N_1\!+\!2$ of energy $1$ and $N_2\!-\!1$ of energy $2$);
\item there are exactly three particles of energy $1$ and no particles of energy $2$ in a given cell:
\begin{equation}\label{pom232R111p}
w_2~=~\frac{\binom{(N_1-1)+V-2}{V-2}\binom{(N_2-1)+V-2}{V-2}} {\binom{(N_1+2)+V-1}{V-1}\binom{(N_2-1)+V-1}{V-1}}
\end{equation}
\item $w_3=A_3$ that the
reaction $ (3,0) \rightarrow (1,1)  $ occurs.
\end{enumerate}
The other terms in the right-hand side of the MEq~(\ref{meq}) are derived in a similar way. Constructing these terms, one needs to remember that the energy conservation requires $N\!+\!1=(N_1\!+\!2)+(N_2\!-\!1)$, and $N\!-\!1=(N_1\!-\!2)+(N_2\!+\!1)$.

\begin{figure}[t]
	\begin{center}
		\includegraphics[width=1\textwidth]{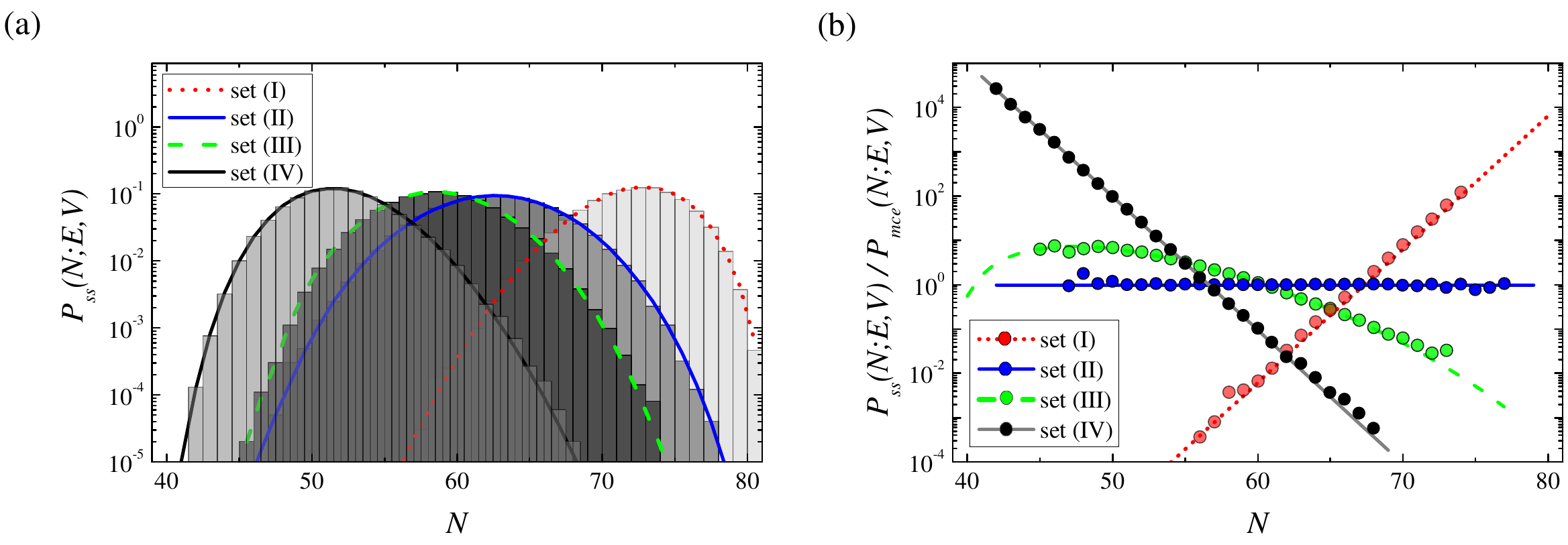}
	\end{center}
	\caption{The SS particle number distributions for $E=80$, $V=20$, and four sets of conditional reaction probabilities: (I) $A_3=0.2,\;A_4=0.3,\;C_3=0.4,\;C_4=0.6$, which satisfy~Eq.~(\ref{gmce}) for $r=2$, (II) $A_3=A_4=C_3=C_4=1$, which satisfy~Eq.~(\ref{CMEUdefPN}), (III) $A_3=0.6,\;A_4=0.1,\;C_3=0.3,\;C_4=0.3$, and (IV) $A_3=0.4,\;A_4=0.6,\;C_3=0.2,\;C_4=0.3$, which satisfy~Eq.~(\ref{gmce}) for $r=0.5$. (a) Comparison of the Monte Carlo (gray histograms) and MEq results~(\ref{232PNU}) (lines). (b) Numerical confirmation of Eq.~(\ref{232balanceU}). Points stand for results of the Monte Carlo simulation, lines represent SS solutions of the MEq.}
	\label{fig:gmce}
\end{figure}
%%%%%%%%%%%%%%%%%%%%%%%%%%%%%%%%%%%%%%%%%%

For fixed values of $E$ and $V$, the SS solution of the MEq~(\ref{meq}) can be obtained from the condition
\begin{equation}\label{meq232e}
\Delta P(N,t)=0
\end{equation}
using recursive relations with the initial condition $P(\frac{E}{2}-1)=0$ (details are provided in Appendix~\ref{PSS}).
The solution reads as follows:
\begin{equation}\label{232PNU}
P_{\rm ss}(N; E,V)~=~\frac{W_{\rm ss}(N;E,V)}{W_{\rm ss}(E,V)}~,
\end{equation}
where, for even values of $E$,
\eq{
\label{232pNU}
& W_{\rm ss}(N;E,V)~=~\binom{N_1+V-1}{V-1}\binom{N_2+V-1}{V-1}~ T(N;E,V)~,\\
\label{TNU1}
& T(N;E,V)~=~\prod_{k=\frac{E}{2}}^{N-1}\frac{(2k\!-\!E)(E\!-\!k\!+\!V\!-\!3)\;C_3
+(E\!-\!k\!-\!1)(2k\!-\!E\!+\!V\!-\!2)\;C_4}
{(2k\!-\!E)(E\!-\!k\!+\!V\!-\!3)\;A_3
+(E\!-\!k\!-\!1)(2k\!-\!E\!+\!V\!-\!2)\;A_4 }~,\\
& W_{\rm ss}(E,V)~=~\sum_{N=\frac{E}{2}}^{E} W_{\rm ss}(N;E,V)~.\label{CEV}
}
where, from the definition of the empty product, $T(\frac{E}{2};E,V)=1$. Let us note that in the pathological case of $C_3=C_4=0$ one has $T(N;E,V)=\delta_{N,\frac{E}{2}}$. In this case, one gets $P_{ss}(N;E,V)=\delta_{N,\frac{E}{2}}$. It can be also shown that the second pathological case,  $A_3=A_4=0$, boils down to $P_{ss}(N;E,V)=\delta_{N,E}$. To show this, one has to reformulate the solution of the MEq by solving it with the initial condition $P(E+1)=0$ (cf. Appendix~\ref{PSS}). For all possible choices of $E$ and $V$, and the conditional reaction probabilities $A_3,\;A_4,\;C_3,\;C_4$, the analytical expression (\ref{232PNU}) for the SS multiplicity distributions agrees with results from Monte Carlo simulations (see Fig.~\ref{fig:gmce}).

When dividing the numerator and denominator of each factor in the product $T(N;E,V)$ (\ref{TNU1}) by an arbitrary constant parameter $\alpha$, the resulting SS distribution does not change. This observation, though evident in itself, allows to see quite nonobvious feature of $P_{\rm ss}(N;E,V)$. Namely, putting, for example, $\alpha=C_3$ one finds that, for fixed $E$ and $V$, $P_{\rm ss}(N;E,V)$ only depends on three independent parameters - the ratios of the  conditional reaction probabilities: $C_4/C_3$, $A_3/C_3$, and $A_4/C_3$, correspondingly.

\begin{figure}[t]
	\begin{center}
		\includegraphics[width=0.6\textwidth]{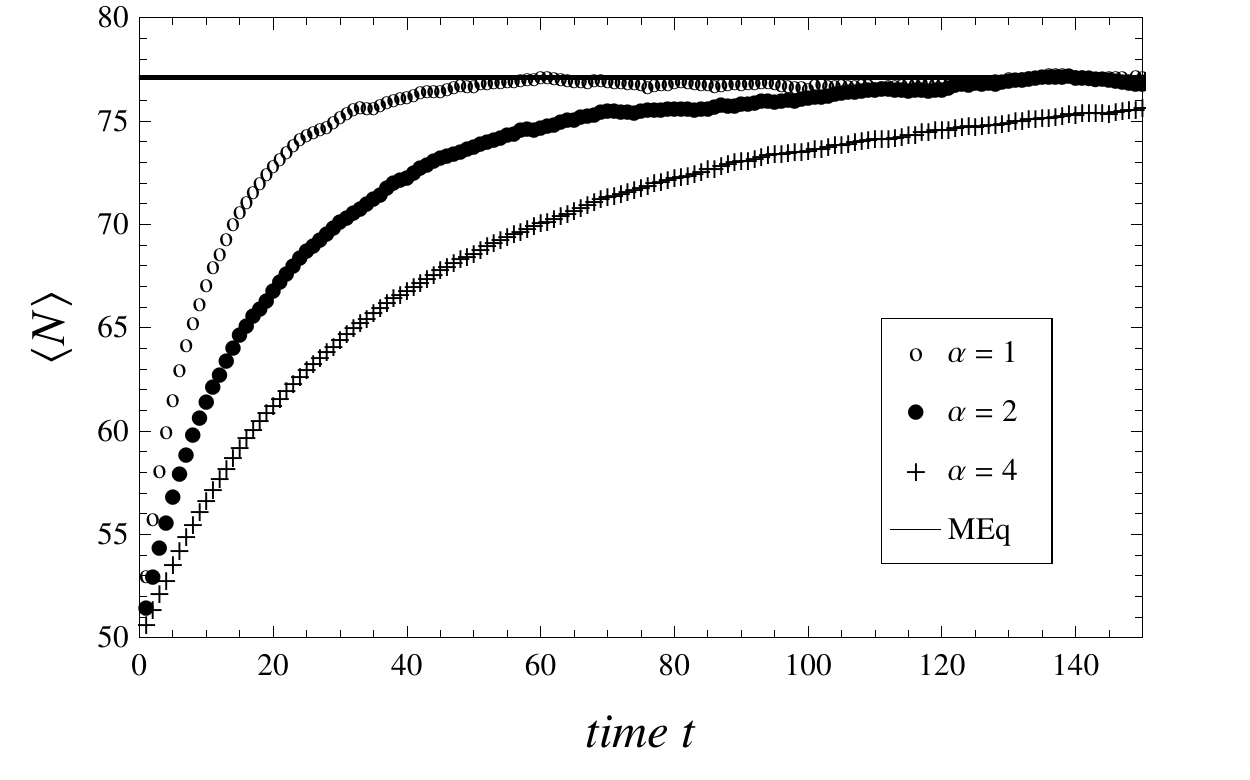}
	\end{center}
	\caption{Mean particle number, $\langle N \rangle$ as a function of the time $t$ calculated for systems with $E=100$, $V=20$, $A_3=A_4=C_3=C_4=1$, and for $\alpha=1$ (circles), $\alpha=2$ (dots), and $\alpha=4$ (crosses). The convergence curves were obtained within Monte Carlo simulations averaged over $1000$ runs. For comparison, the SS result for $\langle N\rangle$ is also shown (solid line).}
	\label{fig:tau}
\end{figure}

Additionally, assuming that $\alpha\geq 1$ one can study the effect of rescaled
conditional reaction probabilities (i.e.~$A_3/\alpha,\;A_4/\alpha,\;C_3/\alpha, \;C_4/\alpha$), on the CM convergence towards the stationary state. This is illustrated in Fig.~\ref{fig:tau}, where the time dependence of the mean particle number is plotted for different values of $\alpha$ and with other model parameters fixed. The results are calculated using the Monte Carlo approach. In the figure, the mean particle number for the SS calculated using the MEq approach is also plotted for a comparison. As expected
Monte Carlo results approach the MEq one with increasing time and the convergence is slower for larger values of $\alpha$.

\section{Steady State Versus Microcanonical Ensemble}
\label{sec:SSvsMCE}

In this section, the particle number distributions calculated for the
SS are compared with the corresponding MCE distributions, and the limits of MCE applicability in dynamical systems are discussed. Firstly, we discuss conditions needed for the agreement between the SS and MCE.
Secondly, these conditions are somewhat relaxed, resulting in the grand microcanonical ensemble (GMCE). Thirdly, the most general case of the SS distribution is considered, that does not correspond to any known distribution of statistical mechanics. Finally, an extreme case of only one type of reactions (creation or annihilation) changing particle number is discussed.

\begin{enumerate}[(i)]
\item One notes that the SS distribution~(\ref{232PNU}) differs from the
MCE one (\ref{CMEUdefPN}) by the factor $T(N;E,V)$, which comprises a dependence on the conditional reaction probabilities. Thus, the SS distribution coincides with the MCE one provided that $T(N;E,V)=1$. This is the case for the special selection of the conditional reaction probabilities, namely,
\eq{
A_3=C_3,~~~~~~A_4=C_4~.
\label{srp_mce}
}
Then the number of independent parameters of the SS distribution is reduced from three to one, which corresponds to the ratio $C_3/C_4=A_3/A_4$. Thus, for any allowed value of this ratio, and for all values of $E$ and $V$, the SS predictions agree with the MCE ones.

The condition~(\ref{srp_mce}) for the equivalence of the SS and MCE was derived within the specific version of the CM discussed in this paper as an example.
Below a more general considerations are presented which are valid for any version
of the Cell Model which is the ergodic Markov chain model (see Appendix~\ref{Markov}).
One considers two microstates,
$X = (n^{\rm X}_{1,1},n^{\rm X}_{2,1};\;\ldots; \;n^{\rm X}_{1,V},n^{\rm X}_{2,V})$ and
$Y = (n^{\rm Y}_{1,1},n^{\rm Y}_{2,1};\;\ldots;\;n^{\rm Y}_{1,V},n^{\rm Y}_{2,V})$,
which can be transformed to each other with conditional probabilities
$B({\rm X}\rightarrow {\rm Y}|\rm X)$ and $B({\rm Y}\rightarrow {\rm X}|\rm Y)$, respectively.
The conditional probability $B({\rm X}\rightarrow {\rm Y}|\rm X)$
is a probability of the transition
${\rm X}\rightarrow {\rm Y}$ given the microstate $\rm X$.
Thus, the transition probability is equal to
$P({\rm X}\rightarrow {\rm Y}) = B({\rm X}\rightarrow {\rm Y}|\rm X) P(X)$
(respectively, $P({\rm Y}\rightarrow {\rm X}) = B({\rm Y}\rightarrow {\rm X}|\rm Y) P(Y)$), where
$P(\rm X)$ is the probability of a microstate $\rm X$.

Let us assume that the conditional  probabilities of forward and backward transitions
are equal for all pairs of microstates,
$B({\rm X}\rightarrow {\rm Y}|\rm X)=B({\rm Y}\rightarrow {\rm X}|\rm Y)$.
Then one first postulates that the starting (at $t=0$) distribution of microstates in a large ensemble of systems is equal to the MCE one, i.e., all microstates have the same probability to appear, $P(\rm X) = P(\rm Y)$.
Consequently, after the first time step ($t=1$) all microstates have the same probability to appear,
as by the assumption, the numbers of forward and backward transitions are equal.
This is, of course, also the case after each subsequent time step. One concludes that the microstate distribution is time independent and, thus, the corresponding SS is equivalent to the MCE. But within the ergodic Markov models (see Appendix~\ref{Markov}),
the SS is independent of a starting distribution
(see Appendix~\ref{Markov}). Therefore, one shows that
equal conditional transition probabilities for ${\rm X} \rightarrow {\rm Y}$ and
${\rm Y}\rightarrow {\rm X}$ transitions result in a SS
which is equal to the MCE. Moreover, the transition probabilities for forward and
backward transitions are equal:
$P({\rm X}\rightarrow {\rm Y}) = P({\rm Y}\rightarrow {\rm X})$.
This corresponds to the so-called detailed balance in the system.

\item Let us relax the condition~(\ref{srp_mce}) by introducing
a parameter, $r$, defined as:
\eq{
C_3 = rA_3,~~~~~C_4=rA_4~.
\label{srpgmce}
}
Then, for fixed $E$ and $V$, the number of independent parameters in $P_{\rm ss}(N;E,V)$ increases from one to two and the factor $T(N;E,V)$ in Eq.~(\ref{TNU1}) reads
\begin{equation}\label{TNU2}
T(N;E,V)~=~r^{N-\frac{E}{2}}~.
\end{equation}
For $r=1$, one gets $T(N;E,V)=1$ and recovers the previously discussed result (\ref{srp_mce}), which coincides with MCE. For $r\neq 1$, the SS particle-number distribution~(\ref{232PNU}) can be written in the form:
\begin{equation}\label{gmce}
P_{\rm ss}(N;E,V)~=~W_{\rm mce}(N;E,V)\frac{\exp(-\mu N)}{Z_{\rm gmce}(E,V,\mu)}
~\equiv~ P_{\rm gmce}(N;E,V,\mu)~,
\end{equation}
where $\mu\equiv -\ln(r)$ resembles a dimensionless chemical potential,
and $Z_{\rm gmce}(E,V,\mu)$ is a normalization factor. $Z_{\rm gmce}(E,V,\mu)$ stands for the partition function of the GMCE. In this ensemble energy is conserved,
whereas particles can be exchanged with a particle bath.
An example distribution of the form~(\ref{gmce}) is presented in Fig.~\ref{fig:gmce}.

One can show that the condition~(\ref{srpgmce}) follows from the assumption of the
balance of transitions between states with $N$ and $N+1$ particles.
Note, that the $N$ and $N+1$ states, in general, include many different microstates.
Thus, the balance condition discussed here (the coarse balance)
is different than the detailed balance
condition discussed previously.
As an example, let us consider the creation reaction
$(1,1) \rightarrow (3,0)$ and the corresponding annihilation reaction.
The other transitions lead to the same conclusions.
The coarse  balance condition for these reactions reads:
\begin{equation}
P_{\rm ss}(N;E,V)\omega_{N\rightarrow N+1}=P_{\rm ss}(N+1;E,V) \omega_{N+1\rightarrow N}~,
\end{equation}
where
$P(N)$ is given by Eq.~(\ref{232PNU}) while the transition probabilities
$\omega_{N\rightarrow N+1}$ and $\omega_{N+1\rightarrow N}$ can be read from the MEq~(14):
\begin{equation}
\omega_{N\rightarrow N+1}=C_3 \frac{\binom{N_1+V-3}{V-2}\binom{N_2+V-3}{V-2}}{\binom{N_1+V-1}{V-1} \binom{N_2+V-1}{V-1}}
\end{equation}
and
\begin{equation}
\omega_{N+1\rightarrow N}=A_3 \frac{\binom{N_1+V-3}{V-2}\binom{N_2+V-3}{V-2}}{\binom{N_1+V+1}{V-1} \binom{N_2+V-2}{V-1}}~.
\end{equation}
After a short algebra one gets
\begin{equation}
\frac{C_3}{A_3}=\frac{T(N+1;E,V)}{T(N;E,V)}~,
\end{equation}
which can be written as
\begin{equation}
\frac{C_3}{A_3}=\frac{f C_3 + g C_4 }{f A_3 + g A_4}~,
\end{equation}
where $f\equiv f(N;E,V)$ and $g\equiv g(N;E,V)$. Finally, by cross-multiplying one gets
\begin{equation}\label{232balanceU}
\frac{C_3}{A_3}=\frac{C_4}{A_4}\equiv r~.
\end{equation}

\item Now, let us consider the most general case of unconstrained conditional reaction
probabilities. As already mentioned, in this case the SS multiplicity distribution depends on three (arbitrary chosen) ratios of the conditional reaction probabilities (e.g., $C_4/C_3$, $A_3/C_3$, and $A_4/C_3$) and as a rule the SS is different from MCE and GMCE and the detailed balance is not fulfilled. Examples of $P_{\rm ss}(N; E,V)$ are presented in~Figs.~\ref{fig1} and~\ref{fig:gmce}.

\item One can also consider an extreme case with only annihilation reactions
taking place, i.e., $A_3>0$, $A_4>0$ and $C_3=C_4=0$. Then all systems approach microstates with a minimum number of particles, $N_{\rm min}= \lceil\frac{E}{2}\rceil$, allowed by the energy conservation, i.e. $P_{\rm ss} (\lceil\frac{E}{2}\rceil;E,V)=1$ and $\forall_{N\neq \lceil\frac{E}{2}\rceil}P_{\rm ss}(N;E,V)=0$. Similarly for the model with only creation reactions taking place, the systems approach microstates with the maximum number of particles, $N_{max}=E$, i.e. $P_{\rm ss}(E;E,V)=1$ and $\forall_{N\neq E}P_{\rm ss}(N;E,V)=0$.
The $P_{\rm ss}(N;E,V)$ distributions for the two extreme settings of the conditional reaction probabilities are different and different than the corresponding MCE distribution. But the detailed balance is fulfilled in the SS.
Namely, there are no inelastic reactions and
probabilities of forward and backward elastic reactions is, by definition, equal.
Thus, one concludes that the detailed balance alone
does not guarantee the agreement between the SS and MCE.
It should be mentioned, however, that the extreme settings of the conditional
reaction probabilities, the CM does not correspond to the ergodic
Markov chain. The ergodicity  is one of the basic properties of statistical mechanics.
\end{enumerate}

Let us note that the presented results are consistent with the maximum entropy formulation of statistical mechanics~\cite{1957Jaynes}. This formulation boils down to maximization of the entropy 
\eq{\label{entropy}
	S~\equiv~ -~\sum_i p_i~\ln p_i~,
}
with respect to the microstate probabilities $p_i$ ( $i$ runs from 1 to the total number of microstates $W_{\rm mce}(E,V)$ )
under given constrains. Then, imposing the constrain of 
$\sum_ip_i=1$, one can derive MCE (~$ p_i = 1/W_{\rm mce}(E,V)~) $  with particle number distributions given by Eq. (\ref{CMEUdefPN}). 
Obviously,  the conditional reaction probabilities
which fulfill Eq.~(\ref{srp_mce}) and lead to to the SS equal to
the MCE also lead to the maximum-entropy SS.
Other selections of the conditional reaction probabilities result in SSs with entropy lower than the maximum one. Thus, with increasing time 
the state entropy may increase, decrease or stay constant. This depends on
the selected state at $t = 0$ and conditional reaction probabilities. 
For example, for the SS at $t = 0$ equal to the MCE and the conditional reaction
probabilities that do not fulfill Eq.~(\ref{srpgmce}) the entropy decreases with time.

By imposing an additional constrain when maximizing Eq.~(\ref{entropy}) 
with respect to $p_i$,
namely, average number of particles is fixed, one finds GMCE with $P_{ss}(N;E,V)$ given by Eq.~(\ref{gmce}) with non-zero chemical potential $\mu$. Finally, for the conditional reaction probabilities that do not fulfill Eq.~(\ref{srpgmce}), the resulting ensemble has a general form that cannot be related to any known ensemble of statistical mechanics.

\section{Closing Remarks}
\label{sec:sum}

Steady states of isolated systems depend on features of reactions between particles. 
The MCE describes isolated systems within statistical mechanics 
and it is independent of reactions between
particles. Thus, properties of an SS and MCE are, in general, different.
This statement seems to be quite evident. 
However, for systems with particle creation and annihilation stochastic reactions, 
we did not find its proof in the literature.

Here, this question is addressed within  the simple stochastic model 
of particle annihilation and creation, the CM. 
The model predicts a particle-number distribution. 
Its time evolution to an SS was calculated using the Monte Carlo numerical 
simulations, whereas an analytical expression for the SS distribution was obtained using the MEq approximation.
This was possible thanks to the simplicity of the model which corresponds to
a discrete Markov chain with only two single particle energies allowed.

Within the model, the SS distribution depends on the system volume and energy 
as well as on three parameters related to annihilation and creation reactions. For almost all sets 
of randomly selected reaction parameters the SS is different 
than the MCE.
%This agrees with the naive expectations. 

The reaction parameters of the Cell Model can be adjusted to grant the agreement between 
the SS and MCE. This takes place when the annihilation and creation conditional 
reaction probabilities are equal. This requirements also leads to the detailed balance in 
the SS.

For the conditional creation probabilities proportional to the annihilation ones with the same proportionality factor for different reactions, one gets the particle number distribution in the SS which disagrees with the MCE one.  The distribution is equal to the distribution for the GMCE with the dimensionless chemical potential equal to logarithm of the proportionality factor. Thus, this case imitates an agreement of the SS with the predictions of statistical mechanics. The requirement of proportionality of the conditional creation and annihilation 
probabilities is derived from the assumption of the balance of transitions between particle number states, the coarse balance.

The results presented here refer to isolated systems, which do not actually
occur in real experiments. Probably the closest to isolated systems are
multi-particle systems created in high-energy collisions between particles or nuclei.
Many features of particle production in these processes resemble predictions of
statistical mechanics. 
Based on the results presented in this paper,  
we conclude that the success of statistical mechanics in modelling  
particle creation in high energy collisions 
is likely to be the consequence of equal  
forward (creation) and backward (annihilation) conditional reaction probabilities.
Alternatively, several models postulate that multi-particle systems are "drawn" from the
MCE~\cite{Hagedorn:1965st,Gazdzicki:1998vd,Castorina:2007eb}.

The requirement of equal creation and annihilation conditional probabilities 
is likely to be related to the time-reversal invariance of microscopic reactions;
the subject is beyond the scope of this paper.
This simple condition is difficult to implement in  modelling of 
microscopic hadron dynamics.
For example,
nucleon - anti-nucleon annihilation leading to production of $n$-pions ($n=2,3,\ldots$) is usually included, but the corresponding backward reactions for $n > 3$ 
are not implemented because of technical difficulties. 
We hope that this work may stimulate  further discussions on
foundations of statistical and dynamical models of particle production
in high-energy collisions. 
The results of this paper can be used for testing the 
relativistic transport models of hadron production in high-energy nuclear physics and
better understanding of the correspondence between these models and the statistical approach. 
Usually, the difference between transport and statistical models in nucleus-nucleus collisions 
are attributed to nonequilibrium effects of transport models. However, one should additionally 
compare both models at the same conditions, i.e.,
at the thermodynamical equilibrium.
%Studies of equilibrated hadronic systems within the transport 
%models can be performed in a box with periodic boundary conditions. 

\begin{acknowledgments}
%%%%%%%%%%%%%%%%%%%%%%%
%

We would like to thank D.V. Anchishkin, St.~Mrowczynski, Yu. Shtanov and Marysia Prior
for fruitful discussions, comments and corrections.
The work of by the Program of Fundamental Research of the Department of Physics and Astronomy of NAS (M.I.G), the National Science Centre of Poland under grants no.~2015/18/E/ST2/00560 (A.F.), no.~2016/21/D/ST2/01983 (M.M.P.)
as well as by the German Research Foundation under grant GA\,1480/2-2 (M.G.).

\end{acknowledgments}

\appendix

%\appendix
\section{Listing of the algorithm used in Numerical Simulations of CM}\label{App1}

Here, we provide listing of the algorithm for simulating CM model. The \texttt{CM} function contains the main loop that iterates over a specified number of time steps of the simulation. At each iteration, the algorithm performs two operations. First, it generates two sequences of cell multiplicities for particles $N_1$ and $N_2$ (lines 35-36 in the Algorithm 1). It can be done with the help of the auxiliary function  \texttt{RANDOMCOMPOSITION} which computes a random composition of $N$ elements into $V$ groups. In the second part of the main loop (lines 37-54) the appropriate reactions take place. Here, one can easily extend the number of possible reactions to adapt the algorithm for the calculation of other models.

\begin{spacing}{1}
	\label{sec:Annex1}
	\begin{algorithm}[H]
		\caption{Algorithm for CM model}
		
		\begin{algorithmic}[1]
			\Require Energy $E$, number of cells $V$, transition rates $A_3,A_4,C_3,C_4$, number of simulation steps $totalsteps$
			\Ensure Number of particles in each step $\mathbf{M}$
			\Function {RandomComposition}{$N,V$} \Comment{adapted from \cite{Nijenhuis1975}}
			\State Let $\mathbf{n} =\left\lbrace n_i\mid i=1,2,...,V\right\rbrace $ be a sequence of occupation numbers
			\State Let $\mathbf{a} =\left\lbrace a_i\mid i=1,2,...,V\right\rbrace $			
			\State $c_1 \gets V-1$
			\State $c_2 \gets N+V-1$
			\State $i\gets 0$
			\While{True}
			\State $i\gets i+1$
			\If{$c_2\cdot random\le c_1$}
			\State $c_1\gets c_1-1$
			\State $a_{V-c_1-1}\gets i$
			\If{$c_1\le 0$}
			\State $break$
			\EndIf
			\EndIf
			\State $c_2\gets c_2-1$
			\EndWhile
			\State $n_1\gets a_1-1$
			\ForEach{$j: 2\le j\le V-1$}
			\State $n_j \gets a_j-a_{j-1}-1$
			\EndFor
			\State $n_V\gets N+V-1-a_{V-1}$
			\State \Return{$\mathbf{n}$}
			\EndFunction
			\State
			\Function {CM}{$E,V,A_3,A_4,C_3,C_4,totalsteps$}
			\State Let $\mathbf{n} =\left\lbrace n_i\mid i=1,2,...,V\right\rbrace $ be a sequence of cell multiplicities for $N_1$ particles
			\State Let $\mathbf{m} =\left\lbrace m_i\mid i=1,2,...,V\right\rbrace $ be a sequence of cell multiplicities for $N_2$ particles
			\State Let $\mathbf{M} =\left\lbrace M_i\mid i=1,2,...,totalsteps\right\rbrace $ be a sequence of particle multiplicities in consecutive steps of simulation
			\State $step\gets 0$		
			\State $N_1\gets E$ \Comment{exemplary initial condition}
			\State $N_2\gets 0$	
			\While{$step\le totalsteps$}
			\State $step\gets step+1$
			\State $\mathbf{n}\gets RandomComposition(N_1,V)$
			\State $\mathbf{m}\gets RandomComposition(N_2,V)$
			\ForEach{$v: 1\le v\le V$}
			\If{$(n_v=3)\wedge(m_v=0)\wedge(random<C_3)$}
			\State $N_1\gets N_1-2$
			\State $N_2\gets N_2+1$
			\EndIf
			\If{$(n_v=2)\wedge(m_v=1)\wedge(random<C_4)$}
			\State $N_1\gets N_1-2$
			\State $N_2\gets N_2+1$
			\EndIf
			\If{$(n_v=1)\wedge(m_v=1)\wedge(random<A_3)$}
			\State $N_1\gets N_1+2$
			\State $N_2\gets N_2-1$
			\EndIf	
			\If{$(n_v=0)\wedge(m_v=2)\wedge(random<A_4)$}
			\State $N_1\gets N_1+2$
			\State $N_2\gets N_2-1$
			\EndIf		
			\EndFor
			\State $M_{step}\gets N_1+N_2$
			\EndWhile
			\State \Return{$\mathbf{M}$}
			\EndFunction
		\end{algorithmic}
		
	\end{algorithm}
\end{spacing}

\section{Monte Carlo Cell model as Markov Chain } \label{Markov}

Here, we show that CM always converges to a unique stationary distribution $P_{\rm ss}(N;E,V)$.

\begin{figure}[t]
	\begin{center}
		\includegraphics[width=0.6\textwidth]{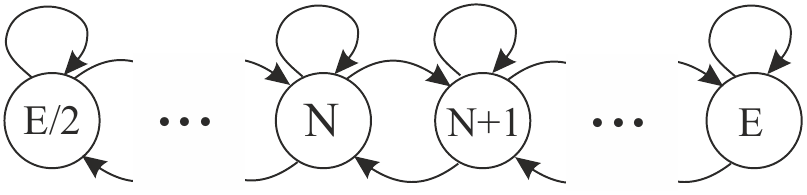}
	\end{center}
	\caption{The spanning subgraph of $\mathcal{G}$.}
	\label{subgraph}
\end{figure}

For a given energy\footnote{In what follows, without loss of generality, we assume that $E$ is even.} $E$, CM is a finite-state Markov chain with states described by the macroscopic quantity $N\in \{\frac{E}{2}, ..., E\}$. Let $\mathcal{M}$ be the transition probability matrix. Elements $\mathcal{M}_{ij}$ of this matrix denote the probability that CM is in the state $j$ at time $t+1$, given that it was observed in the state $i$ at time $t$. Then, one can associate with $\mathcal{M}$ a digraph $\mathcal{G}$ in which each node corresponds to a state of $\mathcal{M}$, and $\mathcal{G}$ contains edge $(i, j)$ if and only if $\mathcal{M}_{ij}>0$~\cite{schier}. In Fig. \ref{subgraph}, we show a spanning subgraph of $\mathcal{G}$ that is further used to prove convergence in distribution of $N$.

In general, to show that any Markov chain converges to a unique stationary distribution, one has to show that it is ergodic~\cite{AMarkov}. In order for the Markov chain to be ergodic, it must satisfy three properties: i) the chain must be irreducible, ii) positive recurrent and iii) aperiodic.
\begin{itemize}
	\item [i)] Since a Markov chain over $\mathcal{G}$ is irreducible iff $\mathcal{G}$ is strongly connected (i.e. there is a path between any pair of nodes $i$, $j$ in $\mathcal{G}$), then it is easy to see from Fig.~\ref{subgraph} that CM is indeed irreducible.
	\item [ii)] CM is recurrent because every irreducible Markov chain with a finite space of states is always recurrent.
	\item [iii)] A sufficient test for an irreducible Markov chain to be aperiodic is that its associated digraph $\mathcal{G}$ contains at least one self-loop (i.e., $\mathcal{M}_{ii}>0$ for some state $i$). This is certainly true for CM (see Fig. \ref{subgraph}), where it is always possible that no reaction in the system occurs in a single time step (i.e. the number of particles does not change).
\end{itemize}

\section{Derivation of $P_{\rm ss}(N;E,V)$ from MEq} \label{PSS}

To solve the MEq (\ref{meq}), one inserts $N_1=2N\!-\!E$ and $N_2=E\!-\!N$ (\ref{CMEdefN1N2}) into this equation. In the stationary regime, when the left-hand side of this equation is equal to $0$ (\ref{meq232e}), its SS solution - $P_{\rm ss}(N;E,V)$ (here for simplicity denoted as $P(N)$) - can be found for even values of $E$ via a recursive scheme starting with the initial condition $P\left(\frac{E}{2}-1\right)=0$. In particular, the first three probabilities can be expressed by the probability $P\left(\frac{E}{2}\right)$ as follows:
\begin{eqnarray}
P\left(1\!+\!\frac{E}{2}\right)\!&\!=\!&\!P\left(\frac{E}{2}\right)
\frac{E V (V\!+\!1)}{2(2V\!+\!E\!-\!2)} \;P\!\left(\frac{E}{2}\right)\;T\left(1\!+\!\frac{E}{2}\right)\!,\nonumber \\
P\left(2\!+\!\frac{E}{2}\right)\!&\!=\!&\!P\left(\frac{E}{2}\right)
\frac{E(E\!-\!2)V(V\!+\!1)(V\!+\!2)(V\!+\!3)}
{24(2V\!+\!E\!-\!2)(2V\!+\!E\!-\!4)}\;T\left(2\!+\!\frac{E}{2}\right)\!,\nonumber\\
P\left(3\!+\!\frac{E}{2}\right)\!&\!=\!&\!P\left(\frac{E}{2}\right)
\frac{E(E\!-\!2)(E\!-\!4)V(V\!+\!1)(V\!+\!2)(V\!+\!3)(V\!+\!4)(V\!+\!5)}
{720(2V\!+\!E\!-\!2)(2V\!+\!E\!-\!4)(2V\!+\!E\!-\!6)}
T\!\left(\!3\!+\!\frac{E}{2}\right)\!,\;\nonumber
\end{eqnarray}
where
\begin{eqnarray}
T\left(1\!+\!\frac{E}{2}\right)&\!=\!&
\frac{C_4}{A_4}~,\nonumber\\
T\left(2\!+\!\frac{E}{2}\right)&\!=\!&T\left(1\!+\!\frac{E}{2}\right)
\frac{2(2V\!+\!E\!-\!8)C_3+(E\!-\!4)VC_4}
{2(2V\!+\!E\!-\!8)A_3+(E\!-\!4)VA_4}~,\nonumber\\
T\left(3\!+\!\frac{E}{2}\right)&\!=\!&T\left(1\!+\!\frac{E}{2}\right)
T\left(2\!+\!\frac{E}{2}\right)\frac{4(2V\!+\!E\!-\!10)C_3+(E\!-\!6) (V\!+\!2)C_4}{4(2V\!+\!E\!-\!10)A_3+(E\!-\!6)(V\!+\!2)A_4}~\nonumber,
\end{eqnarray}
and, in general,
\begin{equation}
P\left(k+\frac{E}{2}\right)=P\left(\frac{E}{2}\right)\;\frac{(V)_{2k}  \left(\frac{E}{2}+V\right)_{-k}}{(2k)!\left(\frac{E}{2}+1\right)_{-k}} \;T\left(k\!+\!\frac{E}{2}\right)~\nonumber,
\end{equation}
where the Pochhammer symbol $(a)_b = a(a + 1)...(a + b-1)$ refers to a rising factorial starting at $a$. By noting that for integer $a$ and $b$
\begin{equation}
(a)_b=\frac{(a+b-1)!}{(a-1)!}=\binom{a+b-1}{a-1}b!~,\nonumber
\end{equation}
one obtains:
\begin{equation}
P\left(k+\frac{E}{2}\right)=P\left(\frac{E}{2}\right)\;\frac{\binom{\frac{E}{2}-k+V-1}{V-1}\binom{2k+V-1} {V-1}}{\binom{\frac{E}{2}+V-1}{V-1}} \;T\!\left(k+\frac{E}{2}\right)~.\nonumber
\end{equation}
Finally, substituting $N=k+\frac{E}{2}$ into the last equation, one gets the solution of MEq, cf.~Eqs.~(\ref{232PNU}) and~(\ref{232pNU}):
\begin{equation}\label{app1}
P(N)=P\left(\frac{E}{2}\right)\;\frac{\binom{E-N+V-1}{V-1} \binom{2N-E+V-1}{V-1}}{\binom{\frac{E}{2}+V-1}{V-1}}\;T(N)~,\nonumber
\end{equation}
where $T(N)\equiv T(N;E,V)$ is given by Eq.~(\ref{TNU1}) and the term  $P\left(\frac{E}{2}\right)$ can be calculated from the normalization condition:
\begin{equation}
\sum_{N=\frac{E}{2}}^{E}P(N)=1~.\nonumber
\end{equation}

The SS distribution $P_{\rm ss}(N;E,V)$ for odd values of $E$ can be found in a similar way, by performing a recursion with the initial condition $P(\frac{E-1}{2})=0$.

%%%%%%%%%%%%%%%%%%%%%%%%%%%%%%%%%%%%%%%%%%

\end{document}